\documentclass[aps,prl,twocolumn,amssymb,showpacs,superscriptaddress]{revtex4}

\usepackage{graphicx}
\usepackage[english]{babel}
\usepackage{bm}    
\usepackage{subeqnarray}

\newcommand{\be}{\begin{equation}}
\newcommand{\ee}{\end{equation}}

\def\spose#1{\hbox to 0pt{#1\hss}}
\def\ltapprox{\mathrel{\spose{\lower 3pt\hbox{$\mathchar"218$}}
 \raise 2.0pt\hbox{$\mathchar"13C$}}}
\def\gtapprox{\mathrel{\spose{\lower 3pt\hbox{$\mathchar"218$}}
 \raise 2.0pt\hbox{$\mathchar"13E$}}}

\begin{document}

\title{Vehicular traffic flow at a intersection controlled by signal light with a new probability\footnote{Supported by the National Natural Science Foundatin of China under Grand No. 10447111}}
\author{ZHANG Wei\footnote{twzhang@jnu.edu.cn} ,
 ZHANG Wei\footnote{tzwphys@jnu.edu.cn, wzhang2007065@gmail.com} }
\address{$^{1}$ Department of Physics,
Ji Nan University, Guangzhou 510632, China}

\date{\today}

\begin{abstract}
We introduced a probability of traffic light, $P_{\rm L}$, at an intersection when approaching cars in two roads are in same conditions. As a
application, we proposed a modified Nagel-Schreckenberg cellular
automata model for describing a conflicting vehicular traffic flow
at the intersection.  The results show that the plateau region in the fundamental diagrams, caused by the effect of interaction, is dependent not only on the probability $P_{\rm L}$, but also on the adaptive schemes. 
\end{abstract}
\pacs{89.40.-a, 02.50.Ey, 05.40.-a, 05.65.+b}
\maketitle

Efficient transportation systems are essential for the every day
activities of modern industrialized societies. The urgency to reduce $\rm CO_2$ emisions and fuel consumption, and the excessive, unpredictable travel times during traffic congestion call for more efficient control approaches, in particular, the optimization of traffic lights in urban road networks.
During past years, various traffic models have been developed to investigate the problems of traffic jams \cite{Chowdhury2000,Nagatani2002}, car accidents \cite{Huang2001,zhang2006,Yang2007}  and energy dissipation \cite{Shi2007,zhang2008}  within the framework of single lane traffic models. 
Recently, a research focus was put on urban road networks, which required extending one-dimensional traffic models in order to cope with situations \cite{Helbing2005,Helbing2006}. Evidently the optimization of traffic flow at a single intersection is  preliminary but important step for not only global optimization in city networks but also local optimization. After the first model for simulation  two crossing roads\cite{Nagatani1993}, controlling traffic flow at intersections have attracted notable attentions\cite{Fouladvand07}.

In principle, the traffic flow at the intersection of two roads can be controlled via two distinctive schemes, with or without signalized traffic light. It is evident that traffic light is unavoidable as the density of cars increases. And it is agreed that a further improvement of the traffic flow requires applying flexible strategies than fixed-time controls. In this letter, we developed a Nagel-Schreckenberg cellular automata model for describing a conflicting vehicular traffic flow at an intersection and present our simulation results for three intelligent controlling schemes. Our control schemes are inspired by the model proposed by Foulaadvand \cite{Fouladvand07}. In that work, approaching car to the intersection yield to traffic at the perpendicular direction by adjusting its velocity to a safe value to avoid collision. And the priority is given to the nearest car to the intersection according to driving rules. A nature and important question is that which one will get the priority when two approaching cars are in same conditions? In this letter, we introduce a probability for traffic light, $P_{\rm L}$, while that happens. In order to capture the basic feather of this problem, we have constructed a modified NS cellular automata model with three traffic responsive schemes applying $P_{\rm L}$.

We first discuss the traffic adaptive controlling scheme in which the
light signalization is adapted to the traffic.
Nowadays advanced traffic control systems anticipate the
traffic approaching intersections. The data obtained via traffic detectors installed at
the intersection make it
possible to measure the velocity  and distance  of an approaching car and  estimate the time
 the car successively passing through the intersection\cite{Fouladvand04}. That supports us to promote three possible
traffic adaptive controlling schemes.

In each scheme, the distance to the intersection, the velocity and the time passing the crossing point  of the approaching cars are denoted by $d_1$,$v_1$,$t_1$ and $d_2$,$v_2$,$t_2$ for the first
and second road, respectively. Generally, both roads are green. By green (red) road we mean the road for which
the traffic light is green (red). If the two approaching cars both can pass the crossing point at the next step, one road should change to red  to avoid collision  under the next three algorithms.

\begin{enumerate}
\item   The movement priority is given to the nearest car($d_{\rm min}$) to the crossing point, and the
car adjusts its velocity as usual with its leading car. In contrast, the other road is turned red and the car brakes irrespective of its direct gap. The simplest way to take into this
cautionary braking is to adjust the gap with the crossing point itself.
If both the approaching cars have the same distance, $d_1=d_2$, the priority is given to one road with probability $P_{\rm L}$, and the other with $1-P_{\rm L}$;

\item The priority is given to the car that use less time($t_{\rm min}$) for pass through.
While $t_1=t_2$, the priority is given to one road with probability $P_{\rm L}$, and the other with $1-P_{\rm L}$.

\item   No matter what the $d_i$,$v_i$,$t_i$ is, randomly chose one road and give it the movement priority with probability $P_{\rm L}$, and the other with $1-P_{\rm L}$.
\end{enumerate}

We do not use the velocity alone to form a new scheme not only because of
the  traffic safety. If we give the priority to the fastest car
($v_{\rm max}$), the approaching car in red road will lost its velocity.
For high densities, at the intersection, the approaching car in the
green road will always be faster than that in the red road, and the
red road will keep in red. With the velocity priority scheme, the
current of the system rises to its maximum value rapidly, and then
exhibits linear decrease versus density the same manner as in the
fundamental diagram of a single road (data not shown).

\begin{figure}
\begin{center}
\includegraphics[scale=0.2]{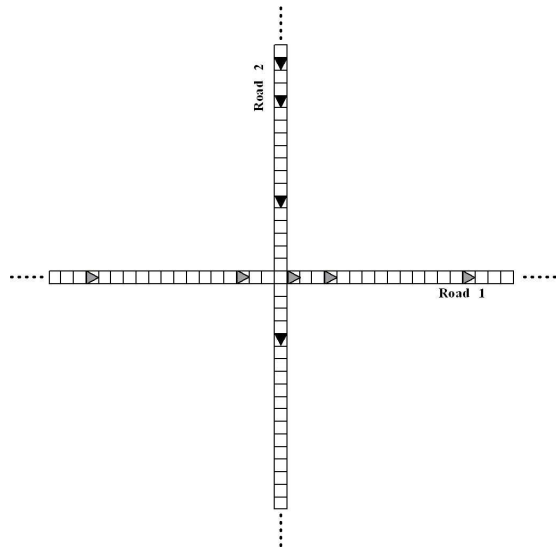}
\end{center}
\caption{ Intersection of two uni-directional roads. They intersect each other at the end.
A closed boundary condition is applied
} \label{fig01}
\end{figure}

Referring to Fig. \ref{fig01}, we consider two perpendicular
one-dimensional chains which represent urban roads accommodating
unidirectional vehicular traffic flow. They cross each other at the
sites $i_1 = i_2 = 0$ on the first and the second chains
respectively. With the closed boundary conditions , the system is
equivalent to cars moving in two circles cross at one point with a
traffic light. The roads, in principle, can each carry two opposite
flows of vehicles. With no loss of generality, we take the direction
of traffic flow in the first chain from north to south and in the
second chain from west to east. Space and time are discretized in
such a way that each chain is divided into cells which are the same
size as a typical car length. Time is assumed to elapse in discrete
steps of $1$s. We take the number of cells to be $L_{1}=L_{2}=500$
for both roads. Each cell can be either occupied by a car or be
empty. Moreover, each car can take discrete-valued velocities $0$,
$1$, $2$, ..., $v_{\rm max}$ in which $v_{\rm max}$ is the maximum
velocity of cars. To be more specific, at each step of time, the
system is characterized by the position and velocity configurations
of cars and the traffic light state at each road.

The system evolves under the NS dynamics
\cite{Nagel92}. Let us briefly explain the NS updating rules which
synchronously evolve the system state from time $t$ to $t + 1$. We
denote position, velocity of the $ith$  car at time step $t$ by
$x(i,t)$ and $v(i,t)$ , respectively. The same quantities for its
leading car are correspondingly denoted by $x(i+1,t)$ and $v(i+1,t)$
. The number of empty cells in front of the $ith$ car is denoted by
$d(i, t) = x(i + 1, t)-x(i, t)-1$. Concerning the above
considerations, the following updating steps evolve
the position and the velocity of each car in parallel.

(1) Acceleration:

 $v(i, t + 1/3)\rightarrow min[v(i, t) + 1, v_{\rm max}]$;

(2)Slowing down:

 $v(i, t + 2/3) \rightarrow min[v(i, t + 1/3), d(i,t)]$;

 (3)Stochastic braking:

 $v(i, t + 1) \rightarrow max[v(i, t + 2/3) - 1, 0]$ with
the probability $p$;

 (4) Movement:

$x(i, t + 1) \rightarrow x(i, t) + v(i, t +1)$.

The state of the system at time $t+1$
is updated from that at time $t$ by applying the modified NS
dynamical rules.
After transients, two
roads maintain steady-state currents, defined as the number of
vehicles passing from a fixed location per a definite time interval,
denoted by $J_1$ , $J_2$ and the average $J_{\rm avg}=(J_1+J_2)/2$. They are
functions of the global densities $\rho_1 = N_1/L_1$ , $\rho_2 =
N_2/L_2$ and $\rho=(\rho_1+\rho_2)/2$, where $N_1$ and $N_2$ are the number of vehicles in the
first and the second road, respectively. We kept the global
density $\rho_1=\rho_2$ and $v_{\rm max}=5$ during the simulations. Figure 2 exhibits
the fundamental diagram of the isolate system, $J_{\rm avg}$ versus
$\rho$, and the difference in three schemes at the fixed $P_{\rm L}=0.5$, $p=0$.

\begin{figure}
\begin{center}
\includegraphics[scale=0.7]{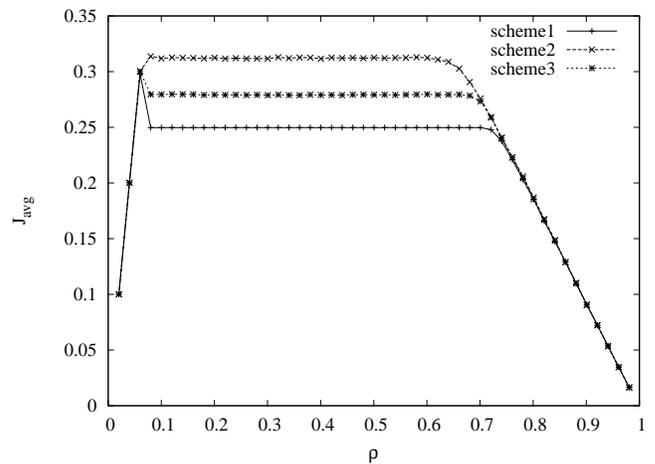}
\end{center}
\caption{$J_{avg}$ versus density at $P_{L}=0.5$ and $p=0$.} \label{fig02}
\end{figure}
It is observed that, in scheme 1 and 3, for small densities
$\rho$, $J_{\rm avg}$ rises to its maximum value rapidly, and then undergoes a short rapid decrease after which a lengthy plateau
region is formed. The current in the plateau region is independent of $\rho$. In scheme 2, $J_{\rm avg}$ rises to the plateau without any decrease. After the plateau, in all three schemes,
$J_{\rm avg}$ exhibits linear decrease versus $\rho$ in the same manner as in the fundamental diagram of a single road. 
It is known that local non-linear interactions can lead to system-wide patterns of motion and 
a local defect can affect the low-dimensional non-equilibrium systems on a global
scale has been confirmed for cellular automata models describing
vehicular traffic flow \cite{Yukawa94}. As discussed in \cite{Fouladvand07}, the intersection makes the cross point as a sitewise dynamical defect whose effect is to form the plateau region in our case. In fact, when $P_{\rm L}=0.5$, the model yield to a non-signaled intersection and the scheme one is equal to that discussed by Foulaadvand. We observed that the $J_{\rm avg}$ in scheme 2 is higher than others. The reason is that, in scheme 2, the car got the priority requires less time to pass through. This raises the efficiency of the intersection compared to the two other schemes hence the second scheme is more optimal than them.
   
\begin{figure}
\begin{center}
\leavevmode
\includegraphics[scale=0.7]{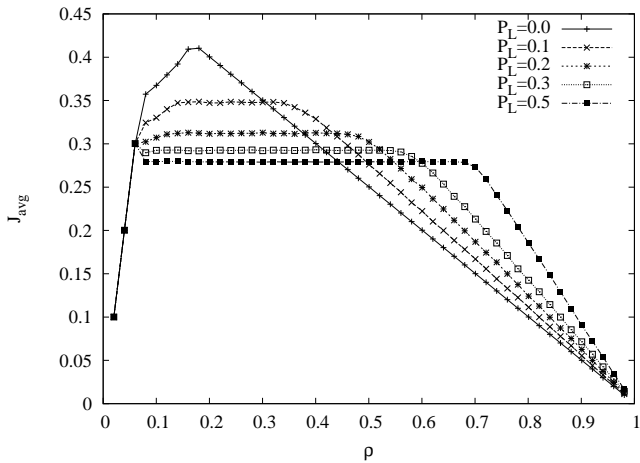}
\end{center}
\caption{$J_{avg}$ versus  $\rho$ at $p=0$ for various values of $P_L$ in scheme3.} \label{fig03}
\end{figure}
\begin{figure}
\begin{center}
\leavevmode
\includegraphics[scale=0.7]{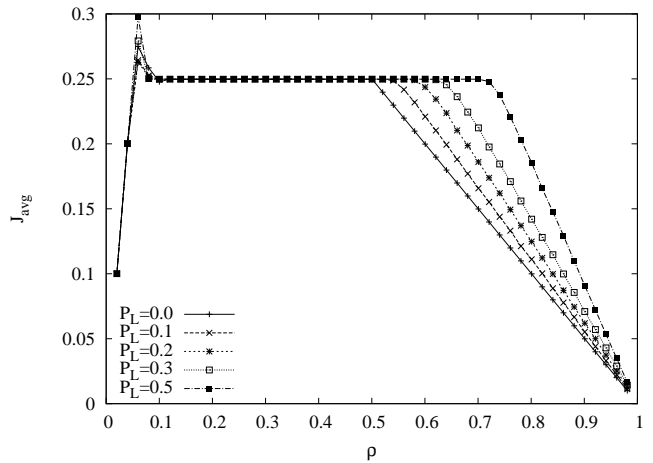}
\end{center}
\caption{$J_{avg}$ versus  $\rho$ at $p=0$ for various values of $P_L$ in scheme1.} \label{fig04}
\end{figure}
Intersection of two chains makes the intersection point appear as a
sitewise dynamical defective site, and the smaller the $P_{\rm L}$ is, the stronger the dynamic defect is. For each value of $P_{\rm L}$, consider the current $J_1$($J_2$) in each road, the larger the $P_{\rm L}$ is,
the plateau region is wider and the current value is more reduced£¨data not shown£©. In order to find deeper insight, it would be illustrative to look at the behavior of average current as a function of density. Fig. \ref{fig03} show that in scheme 3, with fixed $p=0$, $\rho_1=\rho_2$ and $0\leq P_{\rm L} \leq 0.5$. The results show that the plateau region of $J_{\rm avg}$ also becomes wider and the value becomes smaller as $P_{\rm L}$ increases. Because of the equivalence of the two roads, the condition inverts while $0.5\leq P_{\rm L}\leq 1$. In particular,
both with $P_{\rm L}=0$ and $P_{\rm L}=1$, the fundamental diagram exhibits the same manner as that of a single road. The results are same as that in scheme 2. But there is little different in scheme 1. As shown in Fig.\ref{fig04}, with fixed $p=0$, $\rho_1=\rho_2$ and $0\leq P_{\rm L}\leq0.5$, while the length of the plateau region grows with the $P_{\rm L}$ increase, the height of the plateau do not show significant dependence on $P_{\rm L}$. 
While for low density of roads, $P_{\rm L}$ should be adapted according the situation for higher $J_{\rm avg}$, for high density, $P_{\rm L}$ of two roads should be set the same, $P_{\rm L}=0.5$, for optimal purpose. While for low density, $P_{\rm L}$ should be adapted according the situation for higher $J_{\rm avg}$

\begin{figure}
\begin{center}
\leavevmode
\includegraphics[scale=0.7]{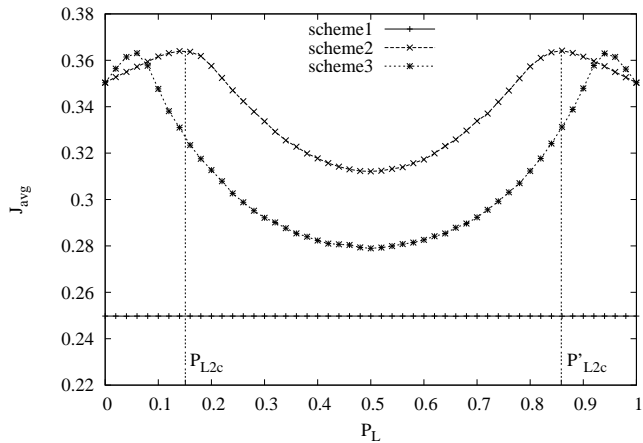}
\end{center}
\caption{$J_{avg}$ versus  $P_{L}$ at $\rho=0.3$ and $p=0$.} \label{fig05}
\end{figure}

We then examined $J_{\rm avg}$ versus $P_{\rm L}$. Because the two roads are equivalent in our model, the symmetry centered at $P_{\rm L}=0.5$ is expected, as shown in Fig \ref{fig05}. In general, the dependence of $J_{\rm avg}$ on $P_{\rm L}$ depends on the value of $\rho$. For large values of $\rho$, $\rho_{\rm c1}>0.70$ in scheme 1 and $\rho_{\rm c2}>0.61$ in scheme 2 and $\rho_{\rm c3}>0.66$ in scheme 3, $J_{\rm avg}$ increases with $P_{\rm L}$, then starts its decrease after the maximum at $P_{\rm L}=0.5$. For $\rho <\rho_{\rm ci}$, the results are different in three schemes. In scheme 2, $J_{\rm avg}$ increases with $P_{\rm L}$ to the maximum at $P_{\rm L2c}$, then the value reduced untill $P_{\rm L}=0.5$. After that point, $J_{\rm avg}$ exhibits an increase up to $P'_{\rm L2c}$ and decrease subsequently.
 We note that the region, $P_{\rm L} \in [P_{\rm L2c},P'_{\rm L2c}]$, is interested. For each value of $\rho$, this region is on the same line, though $P_{\rm L2c}$ and $P'_{\rm L2c}$ are different.
That is the same as shown in scheme 3.
 In scheme 1, there is little different that, for each $\rho$ except large values, $J_{\rm avg}$ is almost independent of $P_{\rm L}$ and remains constant in $P_{\rm L} \in [P_{\rm L2c},P'_{\rm L2c}]$. 
In fact, $P_{\rm L} \in [P_{\rm L2c},P'_{\rm L2c}]$ denotes the appearance of the plateau region as shown in Fig.\ref{fig02}. The results show that, for each $\rho$, the height of the plateau region is independent on $\rho$. In particular, that is also independent of $P_{\rm L}$ in scheme 1.

As shown in Fig.\ref{fig05}, at fixed $\rho=0.3$ and brake possibility $p=0$, while $J_{\rm avg}$ changes little in scheme 1, $J_{\rm avg}$ rises to its maximum value up to $0.363$ and $0.364$ in scheme 2 and 3 respectively. Then $J_{\rm avg}$ reduces rapidly until $P_{\rm L}=0.5$, and the condition inverts subsequently. In scheme 1, the most commonly used method in non-signal intersections, the value of $J_{\rm avg}$ is less than that in  others and insensitive to $P_{\rm L}$. It is easy to see that the scheme 2 is the optimal one. While it is difficult for drivers to get the
velocity and distance to the cross point of another approaching car, the traffic light with
traffic detectors is unavoidable even at $P_{\rm L}=0.5$.

In summary, we introduced a probability of traffic light, $P_{\rm L}$, while cars upon reaching
the intersection are in same conditions. We have investigated the flow characteristics in a signalized intersection via developing a NS model applying $P_{\rm L}$. We have considered three types of controlling schemes. In particular, we have obtained the fundamental diagrams and the dependence of average currant on the probability $P_{\rm L}$road densities.  Our findings show the probability of priority at
the intersection gives rise to formation of plateau regions in the fundamental diagrams and an optimal method for controlling the traffic is unavoidable. The findings may shed light on the way to further study the urban cross road.


\begin{thebibliography}{100}

\bibitem{Chowdhury2000}Chowdhury D, Santen L, and Schadschneider A {\it Phys. Rep.}{\bf 329} 199
\bibitem{Nagatani2002} Nagatani T  {\it Rep. Prog. Phys.}{\bf 65} 1331
\bibitem{Huang2001}Huang D W and  Wu Y P {\it Phys. Rev. E} {\bf 63} 22301-1
\bibitem{zhang2006}Zhang W, Yang X Q, Sun D P, Qiu K and Xia H {\it J. Phys. A: Math. Gen} {\bf 39} 9127
\bibitem{Yang2007}Yang X Q, Zhang W, Qiu K, Xu W T, Tang G, Ren L, {\it Physica A} {\bf 384} 589
\bibitem{Shi2007}Shi W and Xue Y {\it Physica A} {\bf 381} 399
\bibitem{zhang2008}Zhang W, Zhang W, Yang X Q {\it Physica A} {\bf 387} 4657

\bibitem{Helbing2005}Helbing D, Jiang R, and  Treiber M {\it Phys. Rev. E} {\bf 72} 046130.
\bibitem{Helbing2006}Helbing D, Johansson A,  Mathiesen J, Jensen M H, and Hansen A  {\it Phys. Rev. Lett.} {\bf 97} 168001
\bibitem{Nagatani1993}Nagatani T {\it J. Phys. A} {\bf 26} 6625
\bibitem{Fouladvand07}
Foulaadvand M E and  Belbasi S 2007 {\it J. phys. A} {\bf 40} 8289
\bibitem{Fouladvand04}
Fouladvand M E, Sadjadi Z and Shaebani M R 2004 {\it Phys. Rev. E} {\bf 70} 046132

\bibitem{Nagel92}
Nagel K and Schreckenberg M 1992 {\it J. Phys. I France} {\bf2} 2221

\bibitem{Yukawa94}Yukawa S, Kikuchi M and Tadaki S 1994 {\it J. Phys. Soc. Japan} {\bf 63} 3609

\end{thebibliography}
\end{document}